\input harvmac.tex
\input epsf.tex
%
\font\tenams=msam10 \font\sevenams=msam7 
\newfam\amsfam 
\textfont\amsfam=\tenams
\scriptfont\amsfam=\sevenams  

\def\hexnumber#1{\ifcase#1 0\or 1\or 2\or 3\or 4\or 5\or 6\or 7\or
8\or 9\or A\or B\or C\or D\or E\or F\fi}

\edef\theamsfam{\hexnumber\amsfam}  

\mathchardef\gtrsim="3\theamsfam 26 
\mathchardef\lesssim="3\theamsfam 2E 

\ifx\epsfbox\UnDeFiNeD\message{(NO epsf.tex, FIGURES WILL BE
IGNORED)}
\def\figin#1{\vskip2in}
\else\message{(FIGURES WILL BE INCLUDED)}\def\figin#1{#1}\fi
\def\ifig#1#2#3{\xdef#1{fig.~\the\figno}
\midinsert{\centerline{\figin{#3}}%
\smallskip\centerline{\vbox{\baselineskip12pt
\advance\hsize by -1truein\noindent{\bf Fig.~\the\figno:} #2}}
\bigskip}\endinsert\global\advance\figno by1}
\noblackbox

\def\frac#1#2{{{#1} \over {#2}}}

%
%

\def\PL{{\it Phys. Lett.\ }}
\def\PR{{\it Phys. Rev.\ }}
\def\PRL{{\it Phys. Rev. Lett.\ }}

\def\JHEP{{\it JHEP \ }}

\lref\AHpheno{Nima Arkani-Hamed, Savas Dimopoulos and G. Dvali,
    \PL {\bf B429} (1998) 263; \PR {\bf D59} (1999) 086004; I. Antoniadis, N. Arkani-Hamed, S. Dimopoulos and G. Dvali, \PL {\bf B436} (1998) 257.}

\lref\AHcrystal{Nima Arkani-Hamed, Savas Dimopoulos, and John
March-Russell, {\it Stabilization of Sub-Millimeter Dimensions:
The New Guise of the Hierarchy Problem},
hep-th/9809124.}

\lref\kaloper{N. Kaloper, {\it Crystal Manyfold Universes in AdS
Space}, \PL {\bf B474} (2000) 269, hep-th/9912125.}   
\lref\Nam{S. Nam, \JHEP {\bf 04} (2000) 002, hep-th/9911237.}
\lref\GW{W.D. Goldberger and M.B. Wise, \PRL {\bf 83} (1999) 4922,
hep-ph/9907447.}
\lref\AB{I. Antoniadis and C. Bachas, \PL {\bf B450} (1999) 83,
hep-th/9812093.}
\lref\AHDMR{N. Arkani-Hamed, S. Dimopoulos and J. March-Russell,
{\it Logarithmic unification from symmetries enhanced in the
submillimeter infrared}, In ``The many faces of the superworld'',
ed. M.A. Shifman, pp. 627-648, hep-th/9908146.}
\lref\Dvali{G. Dvali, \PL {\bf B459} (1999) 489, hep-ph/9905204.}
\lref\AHSW{N. Arkani-Hamed, L. Hall, D. Smith, and N. Weiner,
\PR {\bf D62} (2000) 105002, hep-ph/9912453.}
\lref\CK{A.G. Cohen and D.B. Kaplan, \PL {\bf B470} (1999) 52,
hep-th/9910132.}

\lref\Gogone{M. Gogberashvili, {\it Hierarchy problem in the shell
universe model}, hep-ph/9812296.}
\lref\Gogtwo{M. Gogberashvili, {\it Europhys. Lett.} {\bf 49} (2000)
396, hep-ph/9812365.}
\lref\RSone{L. Randall and R. Sundrum, \PRL {\bf 83} (1999) 3370,
hep-ph/9905221.}
\lref\RStwo{L. Randall and R. Sundrum, \PRL {\bf 83} (1999) 4690,
hep-ph/9906064.}

\lref\eotwash{ G.L. Smith, C.D. Hoyle, J.H. Gundlach, E.G. Adelberger, B.R. Heckel, and H.E. Swanson, {\it Short-range Tests of the Equivalence Principle }, \PR {\bf D61} 022001 (2000);
 C.D. Hoyle, U. Schmidt, B.R. Heckel, E.G. Adelberger, J.H. Gundlach,
D.J. Kapner and H.E. Swanson, {\it Submillimeter Tests Of The
Gravitational Inverse Square Law: A Search For 'large' Extra
Dimensions}, hep-ph/0011014. }
\lref\landsberg{G. Landsberg, {\it Mini-Review on Extra Dimensions},
hep-ex/0009038.}
\lref\axion{J. Preskill, M.B. Wise and F. Wilczek, \PL {\bf 120B}
(1983) 127; L.F. Abbott and P. Sikivie, \PL {\bf 120B} (1983) 133; M. Dine
and W. Fischler, \PL {\bf 120B} (1983) 137.}
\lref\witten{E. Witten, {\it D-Branes And K-Theory}, \JHEP {\bf 9812} (1998)
019,  hep-th/9810188.}
\lref\MTW{See for example, C.W. Misner, K.S. Thorne and J.A. Wheeler,
{\it Gravitation} (Freeman, San Francisco). }
\lref\mermin{N.D. Mermin, {\it Crystalline Order in Two Dimensions,} \PR
{\bf 176} (1968) 250.}
\lref\kittel{See for example, C. Kittel, {\it Introduction to Solid State Physics},
John Wiley and Sons, Inc.}
\lref\kronig{R. de L. Kronig and W.G. Penney, {\it
Proc. Roy. Soc. (London)}, {\bf A130} (1931) 499.}

\Title{ \vbox{\baselineskip12pt\hbox{BROWN-HET-1255}
}}
{\vbox{
\centerline{Solving the Hierarchy Problem with Brane Crystals}}}

\centerline{Steven Corley and David A. Lowe}
\medskip

\centerline{Department of Physics}
\centerline{Brown University}
\centerline{Providence, RI 02912, USA}
\centerline{\tt scorley, lowe@het.brown.edu}
\bigskip

\centerline{\bf{Abstract}}

The brane world scenario advocated by Arkani-Hamed et al. transmutes
the hierarchy problem into explaining why extra dimensions have sizes
much larger than the fundamental scale. In this paper we discuss
possible solutions to this problem by considering the compactified
dimensions to be populated by a large number of branes in a crystal
lattice. The experimental consequences of this scenario are
described, including the presence of large energy gaps in the spectrum of
Kaluza-Klein modes.

\vfill
\Date{\vbox{\hbox{\sl January, 2001}}}

\newsec{Introduction}

Around two years ago \AHpheno\ pointed out that it was consistent
with known experiments for extra dimensions to exist with sizes of
order a millimeter. The motivation for this observation came from
string theory, where additional ``curled up'' dimensions are required
for the consistency of the theory. String theory also allows for the
existence of D-branes, which give a way to restrict the Standard
Model fields to a three dimensional slice of the higher dimensional
space. Without this additional entrapment of the Standard Model fields
to a brane, large extra dimensions would be in immediate contradiction
with atomic physics. 

The hierarchy problem becomes re-expressed as
explaining why the size of the extra dimensions $r_0$ is much larger
than the fundamental length scale. Denoting the fundamental scale by
$M_*$ (which we can take to be of order $1$ TeV), and the 
four-dimensional Planck scale
$M_{Pl}=10^{19}$ GeV, one finds
\eqn\hierarchy{
M_{Pl}^2 = r_0^n M_*^{n+2}~,
}
for $n$ flat extra dimensions with size $r_0$. 
 For $n=1$, $r_0$ is required to be of cosmological scales,
which is immediately ruled out. The $n=2$ case requires $r_0$ of order
one millimeter. There remains a hierarchy between $r_0$ and 
$1/M_* \approx 10^{-19}$m.

Current rounds of experiments \eotwash,
\landsberg\ are beginning to place direct constraints on the simplest
scenarios.
Short-range gravity experiments now probe down to $200$ $\mu$m 
strongly constraining the $n=2$ case. Accelerator experiments
constrain $M_* \gtrsim 1$ TeV \landsberg. The strongest constraints arise
from astrophysical considerations. Production of Kaluza-Klein modes in 
supernova SN 1987A places a bound $M_* \gtrsim 30$ TeV for $n=2$ \AHpheno.

One of the main theoretical challenges in
implementing some of the proposed large
extra dimension scenario's is how to stabilize the size of the
extra dimensions, without introducing additional fine tuning problems.
This problem should be more readily addressed in the large extra
dimension scenario, versus the traditional approach of
compactification at the fundamental scale, because the analysis may be
carried out at the classical level.
A variety of solutions have been proposed \refs{\GW \AB \AHDMR
\Dvali \AHSW {--}\CK} in the warped
compactification scenarios, where by warped we mean not only
the Randall-Sundrum warped metric scenarios \refs{\RSone \RStwo
\Gogone{--} \Gogtwo} but also cases
where some other field, eg., a scalar as in \AHSW, has a non-trivial
profile in the extra dimensions.  

Another proposed solution in the
unwarped compactification context of \AHpheno\ is to have a large
number of branes $N$ (one of which would be our world) interacting
in such a way that they form a crystal lattice in the internal
dimensions \AHcrystal.  The interbrane
separation could be the size of the fundamental length scale, 
but the size of the extra dimensions
would nevertheless be large enough to solve the hierarchy problem
for a large enough number of branes. While it is still not
clear how to realize such a brane lattice crystal from a more
fundamental theory such as string theory, these models are
nevertheless of interest from a phenomenological point of view,
especially in light of the fact that they could be tested
experimentally in the near future. 

One is still left with the
problem of explaining the large integer $N$. We take the point of view
that replacing a fine tuned continuous parameter by a large integer
parameter is an improvement, as one can set the integer $N$ once and
for all using
initial conditions, and under suitable circumstances, this integer will
be stable with respect to time evolution. Since the hierarchy is set
by a conserved number $N$, it is automatically stable with respect to
radiative corrections.

In the following we begin by discussing the approach to brane
crystals of Arkani-Hamed et al. \AHcrystal. 
They propose a number of scenarios for stabilizing the extra
dimensions. One particularly compelling example does this
 without introducing additional fine tuning, aside from the
usual problem with the four-dimensional cosmological constant. A
problem with this scenario is the presence of unbalanced charge on a
compact space. Balancing the charge leads us to consider a scenario
where neutral non-BPS branes interact via a nearest neighbor
potential. We show this does lead to a natural solution of the
hierarchy problem. The crystal potential leads to a distinctive
experimental signature for this scenario -- the existence
of large energy gaps in the Kaluza-Klein spectrum.
Such energy gaps have also been noticed in the Randall-Sundrum
scenario brane lattices discussed in \refs{\Nam, \kaloper}.

\newsec{Brane Crystal Review}

We begin by reviewing the model considered in \AHcrystal.
They consider a 3-brane embedded in a universe with 3 large
spatial dimensions and $n$ small spatial dimensions.  The system
is described by a bulk action
\eqn\bulkaction{
S_{bulk} = - \int d^{4+n}x \sqrt{- \det g_{4+n}} \left(M^{2+n}_{*} {\cal R}
+ \Lambda - {\cal L}_{matter} + \cdots \right)
}
and a brane action
\eqn\action{
S_{brane} = -  \int d^{4}x \sqrt{- \det g_4} (f^{4} +
\cdots )~,
}
where ${\cal L}_{matter}$ is the Lagrangian of the bulk matter
fields and the ellipses denote higher derivative terms which
may be dropped at low enough energies. Here $g_4$ denotes the induced
metric on the brane, $g_{4+n}$ denotes the bulk metric, $\Lambda$ is
the bulk cosmological constant and $f^4$ is the brane tension.
Interaction terms between the branes are not included. In the next
section we discuss scenarios where such interaction terms are
relevant.

The metric is assumed
to take the form
\eqn\metric{
ds^2 = \left({r_0 \over r}\right)^{n} (dt^2 - R^2 g_{ij} dx^i dx^j) -
r^2 g_{IJ} dx^I dx^J~,
}
where $R=R(t)$ is the scale factor of the three large dimensions and
$r=r(t)$ that of the $n$ small dimensions, with $r_0=r(0)$.
\foot{\AHcrystal\ do not include the conformal factor
$r^{-n}$ multiplying the large dimensions.  We find it more convenient
however to include it as it will lead to diagonalized kinetic 
terms for $r$ and $R$ in the action below and is the ordinary
conformal factor appearing in Kaluza-Klein reductions.}.
Also we use lower case Latin indices $i,j,\cdots$ to denote the 3 large
spatial dimensions and upper case Latin indices $I,J,\cdots$ for
the $n$ small spatial dimensions.

Inserting this form of the metric into the bulk \bulkaction,
and brane \action\ actions and integrating over the spatial
coordinates results in
\eqn\totaction{
S = 
\int dt {R}^3  \left( M^{n+2}_* r_0^n \left(
- 6 \left(\frac{{\dot R}}{{R}}\right)^2
+ \frac{n(n+2)}{2} \left(\frac{\dot r}{r}\right)^2\right) - \left({r_0
\over r}\right)^{2n} V_{tot}(r)\right)~,
}
after an integration by parts.  In obtaining this result we have
assumed that $\int d^3 x \sqrt{\det g_{IJ}} =1$ and $\int d^n x
\sqrt{\det g_{ij}} = 1$.  The potential $V_{tot}(r)$ is given by
\eqn\potential{
V_{tot}(r) = \Lambda r^n - \kappa n(n-1) M_{*}^{n+2} r^{n-2} 
+ f^4~,
}
where the term proportional to $\kappa$ arises from the curvature
of the $n$ dimensional space. For an $n$-sphere $\kappa=1$
and for an $n$-torus $\kappa = 0$. A similar term could be
added for the three large spatial dimensions, but for large $R$ would
be negligible so we drop it.  

For static solutions the equations of motion are simply given by
\eqn\eom{
V_{tot}(r_0)  =  0~,  \qquad
V^{\prime}_{tot}(r_0) =  0 ~,
} 
assuming that $R(t) = R_0$ is constant.  Note that with the potential
\potential\ $r_0$ is fixed entirely in terms of the bulk interaction.
For the above potential
\potential\ these equations constrain $\Lambda$
 assuming that $r_0$ is chosen to solve the
hierarchy problem \hierarchy.  Specifically let's
assume that the brane tension is set by the
higher dimensional fundamental scale $M_*$ so that
$f^4 \approx  M^{4}_*$.  From \eom, assuming the internal space is an
n-sphere, it follows that
\eqn\cosapprox{
\Lambda \approx \frac{M^{n+2}_*}{r^{2}_0} \approx
M^{n+4}_{*} \left(\frac{M_*}{M_{Pl}}\right)^{4/n}~,
}
where we have used \hierarchy.

For $N$ branes with equal tensions, the $f^4$ term in \potential\ is
replaced by $N f^4$. Solving the $V_{tot}(r_0)=0$ equation leads to
\eqn\Napprox{
N \approx \left(\frac{M_{Pl}}{M_*}\right)^{2(n-2)/n}~.
}
This varies from $1$ for $n=2$ to $10^{20}$ for $n=6$. Note one is
still left with an extra fine tuning problem, in order that the bulk
cosmological constant satisfy the relation \cosapprox. Also note the
brane number $N$ plays no role in fixing the size of the extra
dimensions -- this is entirely determined by fine tuning
$\Lambda$. $N$ is fixed only by requiring the four dimensional
cosmological constant vanish.

For a single brane, 
stability of this solution
follows from the condition $V^{\prime \prime}_{tot}(r_0) > 0$ where
$r_0$ solves the equations of motion \eom. 
This is straightforward to see by expanding the action to
quadratic order in the small perturbation $\delta r$ where
$r = r_0 + \delta r$ and then rescaling $\delta r$ so that
the kinetic term is canonically normalized.
The mass can then be read from the action and is given by
\eqn\radionmass{
m^{2}_{r} = \frac{1}{n(n+2)} 
\frac{r_{0}^2 V_{tot}^{\prime \prime}(r_0)}{M_{Pl}^2}~.
}
Evaluating this for \potential\ gives $m_r \approx 1/r_0$. 
Note the static equations of motion \eom\ 
and stability condition do not depend on the specific form of
$V_{tot}(r)$, and yield strong constraints on the parameters of
more general potentials.

This analysis presumes that derivative couplings of the radion to
higher spin Kaluza-Klein modes may be neglected, which is not true in
general. However this should not change the qualitative conclusions.
The analysis also neglects the Hamiltonian constraint, which
would take the form
\eqn\hamil{
M^{n+2}_* r_0^n \left(
- 6 \left(\frac{{\dot R}}{{R}}\right)^2
+ \frac{n(n+2)}{2} \left(\frac{\dot r}{r}\right)^2\right) + \left({r_0
\over r}\right)^{2n} V_{tot}(r) =0~.
}
This tells us if we really considered perturbations independent of the 
three noncompact spatial dimensions, we would generate an non-zero
energy density everywhere in space, which would lead to expansion or
contraction of $R$. This is easily remedied by generalizing to
perturbations localized in the noncompact spatial dimensions.

\newsec{Interbrane Forces}

We now consider in more detail the effect of an interbrane potential
on the above analysis. We continue to work in an approximation where
the compactification is not warped, i.e. the $4+n$-dimensional metric
does not depend on the internal coordinates $x^I$. This presumes the brane
separation will be stabilized at a size parametrically larger
than the fundamental length scale $1/M_*$, so that treating gravity at
the classical level is sufficient. We also continue to treat
perturbations homogeneous in the spatial directions, with only time
dependence, with the metric ansatz \metric.

Let us generalize the brane action to $N$
branes, with a Born-Infeld type action
\eqn\biac{
S_{brane} = - \sum_{k=1}^N \int d^4x  \sqrt{{\rm
det}\left(g_{\mu \nu} - g_{IJ}\partial_\mu X^{I}_{(k)} \partial_\nu
X^{J}_{(k)}\right)} \left( f_k^4  +\sum_{l<k}
V_{brane}\left(r|X^{(l)}-X^{(k)}|\right) \right)~,
}
with an interbrane potential $V_{brane}$.
Here we have assumed the
metric is 
$g=g(t)$ induced on each brane is the same, and we use
$X_{(k)}=X_{(k)}(t)$ to
denote worldvolume fields corresponding
to the brane positions in the transverse space. 
Indices $\mu, \nu$
label the coordinates $(t,x^i)$. The brane potential has been chosen
to depend on the proper distance separating the branes. This leads to
the factor of $r$ in potential term.  In writing the
Born-Infeld action we have used the worldvolume diffeomorphism
invariance to fix the gauge $X_{(k)}^{\mu} = x^{\mu}$. 
Note that had we included spatial dependence of the fields, 
the brane
potential would take a complicated non-local form, that is difficult
to write down explicitly. A useful analogy for the branes interacting
with the bulk gravitational field is gravitational waves
interacting with a resonant gravitational wave detector, \MTW, as we
will comment further below.

A natural candidate for $V_{brane}$ is a simple Coulomb coupling. This
leads to the most interesting brane crystal
scenario studied in \AHcrystal\ with``non-extensive bulk cosmological
constant'', 
where the hierarchy problem was solved without
the additional fine tuning associated with the bulk cosmological
constant. The radion was stabilized in the infra-red using a negative
curvature term in the internal space, and the Coulomb force was used
to provide a short distance stabilizing force. The difficulty with
this picture is that a collection of like charged branes 
on a compact space carries infinite vacuum energy, since the electric
flux has nowhere to end. 

To remedy this, one could consider brane configurations with zero charge per
unit cell. Of course, once branes with opposite sign are present there
will be attractive forces. For supersymmetric D-branes in string
theory, oppositely charged branes will annihilate.  Furthermore,
to our knowledge, there are no known neutral and stable branes.  

The description of D-brane charges as living in K-theory
groups \witten\ however predicts the existence of
stable non-BPS branes which carry charge in a finite,
or torsion, group.  This charge is not associated with a gauge
symmetry and there is therefore no Gauss law preventing us from
considering $N$ such branes on a compact space.  
It remains an open question as to whether such branes
could be used to construct a stable lattice configuration.

For the moment we take a phenomenological point of view and
assume that a stable lattice can be constructed.
These objects will then not experience a Coulomb
interaction. A Van der Waals interaction is one natural interaction
between such objects, arising from the interaction of induced electric
dipole moments, falling off like $1/r^{2n}$. To obtain a stabilizing
potential, this must be combined with a hard-core repulsive
interaction. Taking our motivation from molecular crystals, a possible
potential would be the $n$-dimensional version of the 
Lennard-Jones
potential 
\eqn\LJpotential{
V_{brane}(rX) = M_*^4 v(M_* rX),\qquad v(x) = 
\left({\beta \over x} \right)^{4n} - \left({\gamma \over x} \right)^{2n}~,
}
where $\beta$ and $\gamma$ are both ${\cal O}(1)$.
One could also imagine an ionic lattice of
branes, with screened Coulomb interactions. An importance difference
with the Coulomb force example is that now $V_{brane}$ will scale like
the number of branes $N$, rather than $N^2$ since nearest neighbor
interactions will be dominant. The precise form of the potential will
not be important for what follows.

We now want to show that the size of the internal space 
is fixed in terms of the number of
branes $N$, rather than by using the bulk quantities $\kappa$ and $\Lambda$,
which generally introduce extra fine tuning problems.  We shall
therefore set $\kappa$ and $\Lambda$ to zero, the former implying
that the compact extra dimensions are flat and for simplicity
we take their geometry to be the torus $(S_{1})^n$.

Consider an interbrane potential of the form
\eqn\vbrane{
V_{brane}(r|X^{(k)}-X^{(l)}|) = M_*^4 v(|X^{(k)}-X^{(l)} |rM_*)~,
} 
where $v(x)$ is not fine tuned. The fundamental scale $M_*$ sets the
scale of the interaction.
We assume $v(x)$ is short ranged, so
only nearest neighbor interactions are dominant.  The static 
equations of motion are then roughly given by
$v(x_0)={\cal O}(1)$ and $v'(x_0)=0$.  These have solutions
$x_0 = {\cal O}(1)$, or more explicitly
\eqn\statsoln{
\Delta X_{} \approx {\alpha \over r M_{*}}~,
}
where $\Delta X_{}$ denotes the coordinate separation
between nearest neighbor branes and $\alpha$ is a constant
of order 10 or so, which we will discuss momentarily.  Using the fact that the
coordinate periodicity
around any of the $S^{1}$'s of our extra dimensions is 1 and
summing up the $\Delta X$'s along one of these dimensions
yields the static value of $r$,
\eqn\rstatic{
r_{0} \approx \alpha {N^{1/n} \over M_{*}}~.
}
We find therefore that the size of the internal dimensions is
set by the number of branes along with the fundamental scale.
The value of $r_{0}$ however was already fixed in \hierarchy\ 
and thus yields the necessary number of branes
\eqn\Nsaturate{
N \approx {1 \over \alpha^{n}} \left(\frac{M_{Pl}}{M_*}\right)^2=
{1 \over \alpha^{n}} 10^{32}~.
}
$\alpha = 1$ corresponds to one brane per fundamental volume in the internal
space, saturating the number of branes.  We have therefore required
that $\alpha$ is of order 10 or larger (but not so large that we have
another fine tuning problem) in order that classical gravity be
a good approximation.

The effective potential induced for $r$ takes the form
\eqn\vtot{
V_{tot}(r) = N f^4 + N V_{brane} (r/N^{1/n})~,
}
where integrating out the brane coordinates sets the coordinate
distance between neighboring branes to $1/N^{1/n}$.
Vanishing of the four-dimensional cosmological constant requires one
fine tuning, corresponding to $V_{tot}=0$, 
but note no additional fine tuning is needed. 
The mass of the radion may be obtained using \radionmass, which gives
\eqn\rmass{
m_r=M_*/\alpha^{n/2}~.
}

We have so far taken vanishing bulk cosmological constant.
This is expected at tree-level if supersymmetry is
unbroken in the bulk. However if supersymmetry is broken on the branes
then
a cosmological constant will appear at one loop and could
change the results.  If the breaking takes place at the
fundamental scale $M_{*}$ on the branes then the induced mass
splittings in the bulk are given by a tree-level gravitational effect 
\AHcrystal\
\eqn\split{\Delta m^{2} \approx N {M^{4}_{*} \over M^{2}_{Pl}}
= {M^{2}_{*} \over \alpha^{n}}~,
}
where the last expression was obtained by evaluating $r_{0}$ at
\rstatic.  By dimensional analysis, this induces a cosmological constant  
$\Lambda_{quantum} \approx (\Delta m^{2})^{(4+n)/2}$
so that the potential gets a contribution of
\eqn\cosmoc{
V_{quantum}(r) \approx {M^{4+n}_* \over \alpha^{(4+n)n/2}}
 r^n \to M^{4}_* N \left({x_0 \over \alpha^{(4+n)/2}} \right)^{n}~,
}
where $x_0$ was defined above.  Therefore at $x_{0} \approx \alpha$ 
the induced cosmological constant contribution to the potential
is subdominant and our original estimates above still apply.  

We expect the brane crystal scenario will only work when the number of
extra dimensions $n \geq 3$. This follows from \mermin\ where it is
shown that classical crystal lattices do not exhibit long-range order
in dimensions two or less. 
For $M_* \approx 1$ TeV, \hierarchy\ places $r_0 \approx 10^{-7}$ m
for $n=3$.

\newsec{Experimental Consequences}

To understand the experimental consequences of our scenario we
must investigate the spectrum of the theory as well as the
couplings to Standard Model fields.   
To do this it is convenient to first fix the
$(4+n)$ dimensional diffeomorphism invariance.  For linearized
perturbations $h_{MN}$ about a flat metric $\tilde g_{MN}$ (with $M,N$
labeling the $4+n$-dimensional space) an infinitesimal
diffeomorphism generated by a vector $\xi_{M}$
acts on the metric as 
\eqn\diffeo{h_{MN} \rightarrow h_{MN} + \nabla_M \xi_N + \nabla_N
\xi_M~,
}
where we have decomposed the metric into a background part
$\tilde{g}_{MN}$ and a fluctuation $h_{MN}$ as 
\eqn\metfluc{g_{MN} = \tilde{g}_{MN} + h_{MN}~.}
On the $X^I_{(k)}$'s this will act as
\eqn\xact{
X^I \to X^I - \xi^I~.}

Part of the gauge symmetry can be fixed by demanding
\eqn\gaugefix{\partial^{M} \bar{h}_{MN} = 0~,}
where
\eqn\Hbar{\bar{h}_{MN} = h_{MN} - {1 \over 2} \tilde{g}_{MN} h ~,}
and indices are being raised and lowered with the background metric.
This gauge choice does not fix the diffeomorphism invariance corresponding
to vectors satisfying the $(4+n)$-dimensional wave equation
$\partial^{M} \partial_{M} \xi_{N} = 0$.  

We begin by expanding the Born-Infeld piece of the brane action for a
general perturbation
\eqn\bipart{\eqalign{
\int d^4 x \sqrt{{\rm det} G_{\mu\nu} } &= \int d^4x
\sqrt{{\rm det}\tilde g_{\mu\nu} } (1+ \half \tilde g^{\mu\nu}
n_{\mu\nu}
+   \cr &  {1\over 8} \left( \tilde g^{\gamma \beta} \tilde g^{\mu\nu} - \tilde
g^{\mu \gamma} \tilde g^{\nu\beta} - \tilde g^{\mu\gamma} \tilde
g^{\nu\beta} \right) n_{\gamma \beta} n_{\mu\nu} +\cdots )~,}
}
where $G_{\mu\nu}$ is the full induced metric on the brane, and
\eqn\nmunu{
n_{\mu\nu} = h_{\mu\nu}+h_{\mu I} {{\partial X^I}\over {\partial
x^\nu}} +h_{I \nu} {{\partial X^I}\over {\partial
x^\mu}} + \tilde g_{IJ}  {{\partial X^I}\over {\partial
x^\mu}} {{\partial X^J}\over {\partial
x^\nu}}~.
}
Since $h_{\mu\nu}$ picks up a nontrivial second-order contribution
under diffeomorphisms  of the form
$\xi^\mu=0$, $\xi^I \neq 0$ 
\eqn\hmunudiff{
h_{\mu\nu} \to h_{\mu\nu} + h_{\mu I} \partial_\nu \xi^I + h_{\nu I}
\partial_\mu \xi^I~,
}
as may be seen by expanding \diffeo,
it is convenient to write $h_{\mu\nu} = j_{\mu\nu}+h_{\mu I}
h_{\nu}^I$, so
$j_{\mu\nu}$ will then be invariant under such
diffeomorphisms. This redefinition also makes clear that
\bipart\ gives rise to a mass term for the four-dimensional 
vector fields $h_{\mu
I}$. A different approach to seeing the vector fields become massive
is discussed in \AHpheno. 
However since we are tuning the four-dimensional
cosmological constant to zero, the overall coefficient of \bipart\
will vanish, and the vectors will be massless (as is the graviton).

To examine terms arising from the brane potential, we again restrict
to perturbations independent of the spatial directions. We gauge fix
the fluctuations of the brane coordinates $X^I_{(k)}$ to zero, as
explained in more detail in appendix A. At quadratic order in the
fluctuations, then only the radion modes $\bar h_{IJ}$ couple to the brane
coordinates $X^I_{(k)}$ in the action.

The modes independent of the internal dimensions will therefore be a
massless graviton $\bar j_{\mu\nu}$, a set of massless vector fields
$\bar h_\mu^I$, and a set of massive radion scalars $\bar h_{IJ}$ with mass given by
\rmass, following through the same calculation.
Each of these modes will be at the bottom of a tower of Kaluza-Klein
states which are standing waves in the internal dimensions. These may
be treated in the same way as Bloch waves \kittel. For the low lying
modes, the effects of interactions may be neglected (at least for
sufficiently large $\alpha$). This is precisely analogous to the case
of gravitational waves propagating through a resonant detector, where
one needs to go to next order in the equations of motion to see the
effect of gravitational interactions on the response of the detector \MTW. 
This gives rise to a typical
Kaluza-Klein spectrum for the spin-two and vector modes $m_k =
|k/r_0|$, while for the radion modes $m^2_k = m_r^2 + (k/r_0)^2$.

It is also interesting to calculate the energy band gap at the edge of
the Brillouin zone boundary, where we have standing waves commensurate
with the lattice spacing of the crystal. In general, interaction
effects will become large there. For the radion modes, we can get a
reasonable estimate of this band gap by taking into account only
the interaction of the radion through the brane potential. For plane
wave modes propagating in the $I$'th direction, the equations of
motion
are the same as that of an electron moving in a periodic array of
delta function potentials. This is a special limit of the
Kronig-Penney model \kronig. The wavefunction takes the form 
\eqn\bloch{
r(x) = e^{i k x} u(x)~,
}
where $x$ is the $x^I$ in question, and $u(x)$ is periodic under
lattice translations. Solving the equations of motion for $u(x)$ a
linear combination of $e^{\pm i Kx}$ yields
\eqn\stuff{
\cos(ka) = \cos(K a) + {1\over 2 \alpha^{n-2}} {\sin( K a) \over Ka}~,
}
where $a$ is the lattice spacing $r_0/N^{1/n}$. Here $Ka$ is to be
identified with $\alpha m/M_*$, where $m$ is the mass of the mode. 
For the first
Brillouin zone boundary $k=\pi/a$. Solving this equation yields the
band gap $\Delta m^2 \sim M_*^2/\alpha^n$. This is the same form as
the mass gap of the radion near $k=0$.

Higher order interactions also lead to a band gap for
the spin-two and vector modes at the Brillouin zone
boundary. Estimating the energy difference between a standing wave
with nodes on the branes versus a standing wave with peaks on the
branes leads to the same calculation as in \split.  We therefore
expect the band gap $\Delta m^2$ to be of roughly the same order of magnitude
as for the radion modes. The picture of the band structure that
emerges is illustrated in figure 1.

\ifig\fone{Illustration of expected band structure in the spectrum of
Kaluza-Klein states. The mass squared is plotted versus the wavevector in the
internal dimension $k$ times the lattice spacing $a\sim \alpha/M_*$. The
orders of magnitude of the scales of the band width and band gaps are indicated.}{\epsfysize=4in\epsfbox{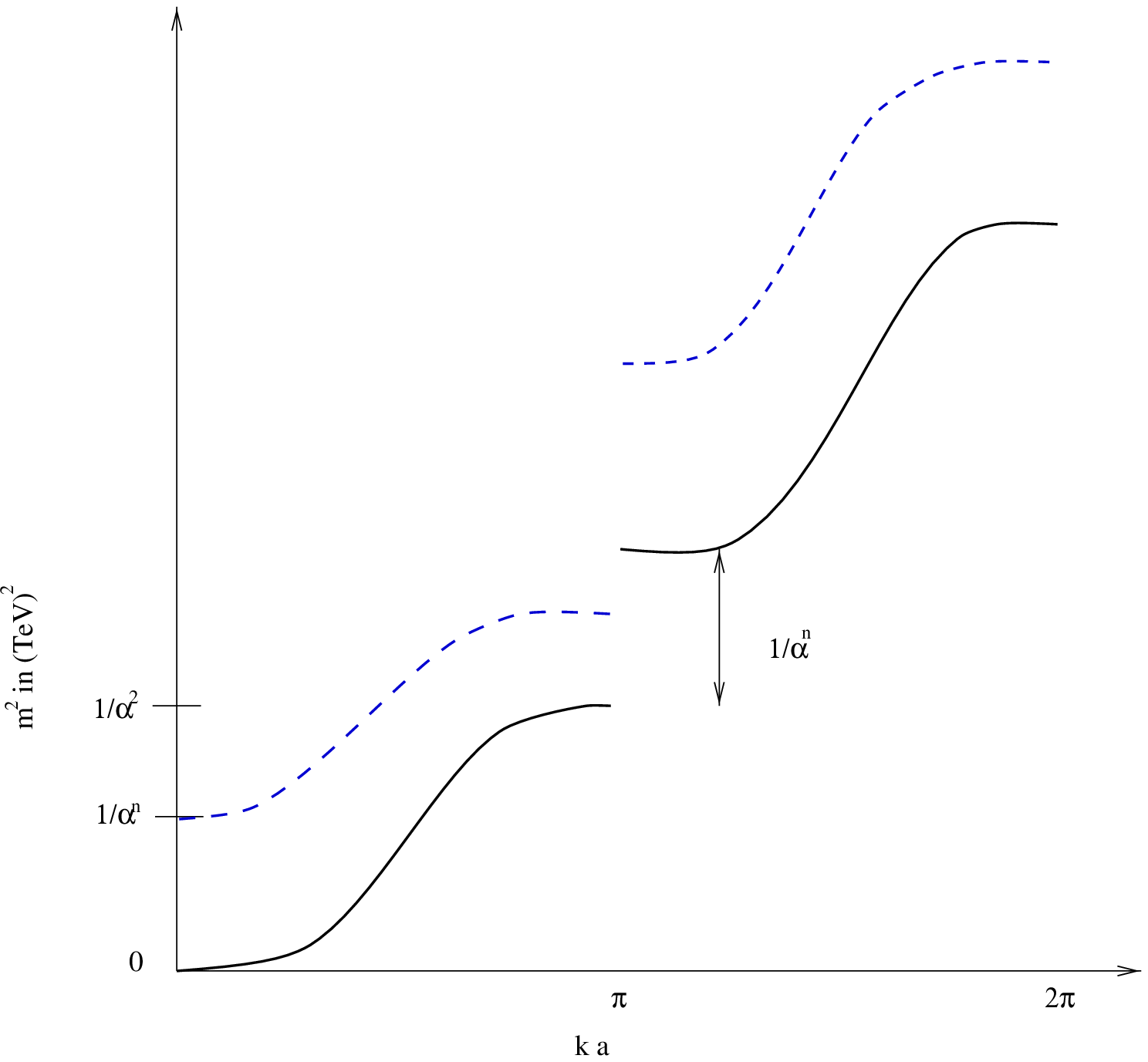}}

The Standard Model fields are coupled to the induced 
metric on the brane via the usual covariant couplings.
Expanding these terms about the background metric
we find that the coupling between the bulk metric fluctuations discussed 
above and the Standard model fields will be suppressed by $1/M_{Pl}$.
The analysis of the phenomenological constraints of \AHpheno\
will therefore carry over to the brane crystal model
unchanged.

\bigskip
{\bf Acknowledgments}
We thank A. Houghton, R. Myers, R. Pelcovits and L. Randall for helpful discussions.
This research is supported in part by DOE
grant DE-FE0291ER40688-Task A.

\appendix{A}{Gauge fixing the brane coordinates}

As noted above, the linearized Einstein equations in covariant gauge
$\partial^M \bar h_{MN}$ have a residual gauge freedom associated with
diffeomorphisms satisfying $\partial^M \partial_M \xi_N =0$.
This freedom can be used
to set the fluctuations of the transverse brane coordinates to zero
given some assumptions.  Specifically the brane coordinates transform
under a diffeomorphism as 
\eqn\Xsym{X^{I}_{(k)}(x^{\mu}) \rightarrow X^{I}_{(k)}(x^{\mu})
- \xi^{I}(x^{\mu}, x^{J})~,}
where we have assumed that $\xi^{\mu} = 0$.  To set the brane coordinate
fluctuation to zero therefore only requires fixing $\xi^{I}$ on 
the brane, and it's extension off the brane can be chosen at
our convenience.  In this case we choose it's extension off the
brane so that it solves the $(4+n)$-dimensional wave equation.
In general this will not be possible.  This follows by noting that
the $\xi^{I}$ live in the bulk spacetime and therefore must satisfy
the periodicity conditions of the extra dimensions.  As a result the
$\xi^{I}$ cannot have arbitrary dependence on the $x^{\mu}$ coordinates
if they 
are to satisfy the wave equation.  To be explicit consider a mode
decomposition of $\xi^{I}$ where one of the modes satisfies 
$\partial^{\mu} \partial_{\mu} \xi^{I} = m^2 \xi^{I}$.  For $\xi^{I}$
to satisfy the $(4+n)$-dimensional wave equation we then need
$\partial^{J} \partial_{J} \xi^{I} = m^2 \xi^{I}$.  For generic
values of $m$ there will be no solution to this equation that
satisfies the required periodicity conditions in the extra dimensions.
Rather the $4$-dimensional mass $m$ will be quantized according
to  $m = \sqrt{q_{1}^2+ \cdots + q_{n}^{2}}/r_{0}$ for arbitrary
integers $q_{i}$.
There is no such constraint on the dependence of the brane coordinates
on $x^{\mu}$ however because they are functions only of the $x^{\mu}$.
We assume for simplicity nevertheless that the brane
coordinate fluctuations can be gauged away in this manner.
In other words we assume that the mode decomposition of the
brane coordinates contains only fluctuations with the quantized
masses given above therefore allowing us to gauge them away.

\appendix{B}{Alternate derivation of linearized equations}

Given these gauge conditions described in appendix A, 
we now give a more general derivation of the linearized equations of
motion describing the metric fluctuations.  We could try to write down 
a potential term describing
the interactions between the branes and then gauge fixing as described above,
but this turns out to be somewhat subtle.  We therefore argue
using symmetry considerations.  The bulk contribution comes
only from the Einstein-Hilbert
term in the action \action.  In the gauge \gaugefix\ it is well
known that this contributes only
$\partial^{P} \partial_{P} \bar{h}_{MN}$ to the linearized equations
of motion.  From the brane terms in the action we
expect the linearized equations of motion to contain
a sum of $\delta$-function terms in the extra
dimensions corresponding to the fixed brane positions with
coefficients determined by symmetry and dimensional analysis.
Specifically we find
\eqn\linear{\eqalign{ 
& \partial^{M} \partial_{M} \bar{h}_{\mu \nu} = 
{M^{4}_{*} \over M_{Pl}^{2}} \sum_{(k)} (a_{(k)}
\bar{h}_{\mu \nu} + b_{(k)} \tilde{g}_{\mu \nu} \bar{h}_{4}) 
\delta^{(n)}(X^{I} - X^{I}_{(k)}) \cr
& \partial^{M} \partial_{M} h_{\mu I} = {M^{4}_{*} \over M_{Pl}^{2}}
 \sum_{(k)} c_{(k)}
h_{\mu I} 
\delta^{(n)}(X^{I} - X^{I}_{(k)}) \cr
& \partial^{M} \partial_{M} \bar{h}_{IJ} = {M^{4}_{*} \over
M_{Pl}^{2}} 
\sum_{(k)} (d_{(k)}
\bar{h}_{IJ} + e_{(k)} \tilde{g}_{IJ} \bar{h}_{n}) 
\delta^{(n)}(X^{I} - X^{I}_{(k)})}}
where $\bar{h}_{4} = \tilde{g}^{\mu \nu} \bar{h}_{\mu \nu}$,
$\bar{h}_{n} = \tilde{g}^{IJ} \bar{h}_{IJ}$, and the coefficients
$a_{(k)},...,e_{(k)}$ are all ${\cal O}(1)$.  The overall factor of
$M^{4}_{*}/M^{2}_{Pl}$ on the right-hand-sides of all equations
is easy to understand by going to a coordinate system $\tilde{X}^{I}
= r_{0} X^{I}$.  In these coordinates $M_{*}$ is the only dimensionful
parameter so that $M^{2-n}_{*}$ would have to be the overall
coefficient following from dimensional analysis.  Going back
to the $X^{I}$ then yields the above coefficient.

Decomposing the fluctuations $\bar{h}_{MN}$ into 
eigenstates of the 4-dimensional wave operator yields
\eqn\decomp{\partial^{M} \partial_{M} \bar{h}_{MN} = 
(- r^{-2}_{0} \partial_{I} \partial_{I} + 
m^{2}) \bar{h}_{MN}}
for a mode with 4-dimensional mass $m$.  It is now straightforward
to estimate the spectrum of metric fluctuations following the
discussion around \bloch.  In particular taking a plane
wave ansatz for the metric fluctuations one recovers the
relation \stuff\ from which we find that the first excited
state and mass gap energies will be or order $m^2, \Delta m^2
\approx M^{2}_{*}/\alpha^{n}$ respectively.

The discussion so far applies for all three equations of motion
in \linear, so in particular it implies that the lowest
energy fluctuation of $h_{\mu \nu}$ would have four-dimensional mass of
order $M_{*}/\alpha^{n/2}$.  This is of course unacceptable if we are to
recover Newtonian gravity on our brane.  For the $h_{\mu \nu}$
equation therefore we must tune the $a_{(k)}$ and $b_{(k)}$
coefficients so that we have a massless fluctuation, or
massless four-dimensional graviton.  This corresponds to fine tuning
the four-dimensional cosmological constant to zero.  Note however
that this fine tuning will not in general imply
that the low lying vector fluctuations $h_{\mu I}$
or radion fluctuations $h_{IJ}$ will be massless.

\listrefs

\end